%% file: ICC2018_final_manuscript.tex
\documentclass[conference]{IEEEtran}
\usepackage{epsfig}
\usepackage{times}
\usepackage{float}
\usepackage{afterpage}
\usepackage{amsmath}
\usepackage{amstext}
\usepackage{amssymb,bm}
\usepackage{latexsym}
\usepackage{color}
\usepackage{graphicx}
\usepackage{amsmath}
\usepackage{amsthm}
\usepackage{graphicx}
\usepackage[center]{caption}
\usepackage{pstricks}
\usepackage{caption}
\usepackage{subcaption}
\usepackage{booktabs}
\usepackage{multicol}
\usepackage{lipsum}
 \usepackage[T1]{fontenc}
\usepackage{epstopdf}

\allowdisplaybreaks

\newtheorem{thm}{Theorem}

\newtheorem{example}{Example}
\theoremstyle{definition}

\providecommand{\definitionname}{Definition}

\newfloat{algorithm}{tbp}{loa}
\providecommand{\algorithmname}{Algorithm}
\floatname{algorithm}{\protect\algorithmname}

\usepackage{changes}
\definechangesauthor[color=red,name={Mingyue Ji}]{MJ}

\input{macros}

\begin{document}

\title{Caching in Combination Networks:
Novel Multicast Message Generation and Delivery by Leveraging the Network Topology}

\author{
\IEEEauthorblockN{%
Kai~Wan\IEEEauthorrefmark{1},
Mingyue~Ji\IEEEauthorrefmark{2},
Pablo~Piantanida\IEEEauthorrefmark{1},
Daniela~Tuninetti\IEEEauthorrefmark{3}
}
\IEEEauthorblockA{\IEEEauthorrefmark{1}L2S CentraleSupélec-CNRS-Université Paris-Sud, Gif-sur-Yvette  91190, France, \{kai.wan, pablo.piantanida\}@l2s.centralesupelec.fr}%
\IEEEauthorblockA{\IEEEauthorrefmark{2}University of Utah, Salt Lake City, UT 84112, USA, mingyue.ji@utah.edu}%
\IEEEauthorblockA{\IEEEauthorrefmark{3}University of Illinois at Chicago, Chicago, IL 60607, USA, danielat@uic.edu}%
}

\maketitle

\begin{abstract}
Maddah-Ali and Niesen's original coded caching scheme for shared-link broadcast networks is now known to be optimal to within a factor two, and has been applied to other types of networks. 
For practical reasons, this paper considers that a server communicates to cache-aided users through $\Hsf$ intermediate relays. In particular, it focuses on {\it combination networks} where each of the $\Ksf = \binom{\Hsf}{\rsf}$ users is connected to a distinct $\rsf$-subsets of relays.
By leveraging the symmetric topology of the network, this paper proposes a novel method to generate multicast messages such that each multicast message sent to each relay is  useful for the largest possible subset of  users connected to this relay. By numerical evaluations, the proposed scheme is shown to reduce the download time  compared to the schemes available in the literature. 
The idea is then extended to decentralized combination networks, more general relay networks, and combination networks with cache-aided relays and users. Also in these cases the proposed scheme outperforms known ones.
\end{abstract}

\section{Introduction}
\label{sec:intro}
{\it Caching} locally popular content is known to reduce network load from servers to users 
during peak traffic times.
A caching scheme comprises two phases. 
(i) {\it Placement phase}: a server places parts of its library into the users' caches without knowledge of later demands. If each user directly stores some bits of the files, the placement is called {\it uncoded}. {\it Centralized} caching systems allow for coordination among users in the placement phase, while {\it decentralized} ones do not. 
(ii) {\it Delivery phase}: each user requests one file. According to the users' demands and cache contents, the server transmits packets to satisfy all the user demands.

In~\cite{dvbt2fundamental} and~\cite{decentralizedcoded}, Maddah-Ali and Niesen proposed caching schemes for  centralized ({cMAN}) and decentralized ({dMAN}) {\it shared-link networks}, respectively, where $\Ksf$ users with cache size $\Msf\Bsf$ bits are connected to a central server with $\Nsf$ files of $\Bsf$ bits through a single error-free shared link. The placement phase is uncoded, while the delivery phase uses linear network coding to deliver carefully designed {\it multicast messages} to the users. The optimality of {cMAN}, in terms of the number of broadcast bits (referred to as {\it load}) for the worst-case set of demands, under the constraint of uncoded placement and $\Nsf \geq \Ksf$ was shown in~\cite{ontheoptimality}, and later extended to $\Nsf < \Ksf$ and to {dMAN} with uniform demands in~\cite{exactrateuncoded}. Uncoded placement is optimal to within a factor $2$ for shared-link networks~\cite{yas2}. 
\begin{figure}
\centerline{\includegraphics[scale=0.16]{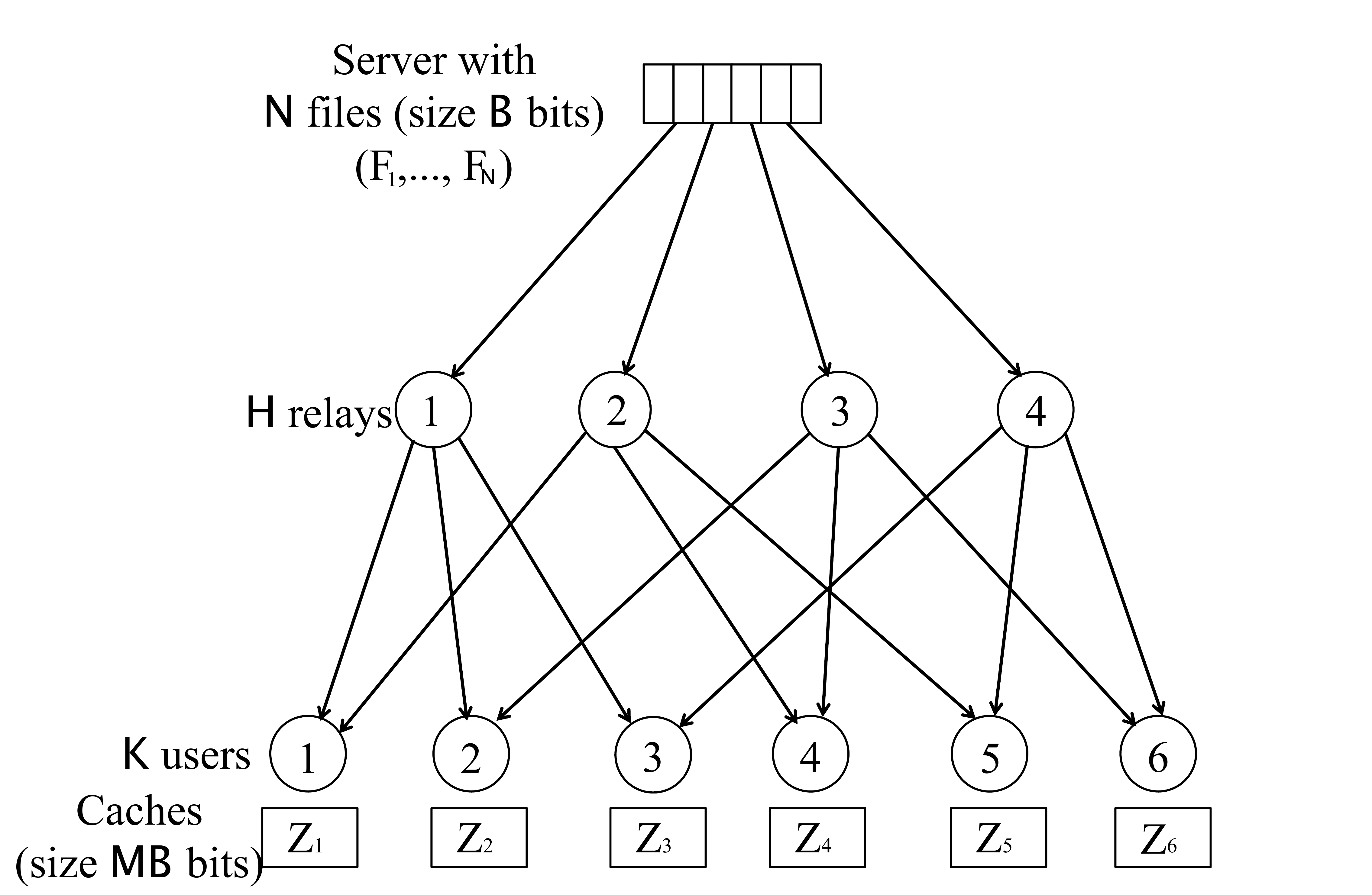}}
\caption{\small A combination network with end-user-caches, with $\Hsf=4$ relays and $\Ksf=6$ users, i.e., $\rsf=2$.}
\label{fig: Combination_Networks}
\vspace{-5mm}
\end{figure}

In practice users and servers may communicate through intermediate relays.  
The caching problem for general relay networks was firstly considered in~\cite{multiserver},  where a scheme based on {cMAN} placement and an interference alignment scheme to transmit {cMAN} multicast messages was proposed. Several works followed, such as~\cite{Karamchandani2016rate, Naderializadeh2017onthoptimality}, but
%
since it is hard to characterize the fundamental limits of 
general relay networks, 
focus has recently shifted to a symmetric network referred to as {\it combination network}~\cite{cachingJi2015}.  As illustrated in Fig.~\ref{fig: Combination_Networks}, in a combination network there are $\Hsf$ relays and $\Ksf= \binom{\Hsf}{\rsf}$ users with cache size of $\Msf\Bsf$ bits, where each user is connected to a different $\rsf$-subset of relays, and all links are error-free and orthogonal.  The objective is to minimize the download time 
for the worst-case demands.
%
The available literature mainly follows the {\it two-step separation principle} formalized in~\cite{Naderializadeh2017onthoptimality}:
(a) {cMAN}-type uncoded cache placement and multicast message generation  (i.e., the generation of the multicast messages is independent of the network topology), and (b) message delivery that aims to match the network multicast capacity region. 
%
Examples of such a separation approach are~\cite{cachingincom,novelwan2017,multiserver}.
In~\cite{cachingincom} two schemes were proposed, one based on routing and the other based on a combination network coding. 
In~\cite{novelwan2017} we proposed a delivery scheme that leverages the structure of the network and then improved upon it for the case $\Msf=\Nsf/\Ksf$ by  interference elimination. 
%
%
Another approach for this problem was proposed in~\cite{Li2016coded,Zewail2017codedcaching}, where the combination network was split into $\Hsf$ shared-link networks and then {cMAN} delivery was used in each one. With a coded cache placement based on MDS (maximum separable distance) codes, the scheme in~\cite{Zewail2017codedcaching} achieves the same performance of~\cite{Li2016coded} but without the constraint in~\cite{Li2016coded} that $\rsf$ divides $\Hsf$.
%
{\it This work departs from these two lines of work and proposes a novel way to generate multicast messages by leveraging the network topology.}

\paragraph*{\textbf{Contributions and Paper Organization}}
We start by considering centralized combination networks. 
In Section~\ref{sec:novel scheme}, we propose a novel delivery scheme which  
generates multicast messages by leveraging the network topology.
Numerical results show that the proposed scheme outperforms existing schemes. 
In Section~\ref{sec:extension}, we extend our novel delivery scheme to
decentralized 
combination networks, 
to general relay networks, and finally
to combination networks with both cache-aided relays and users.

\section{System Model and Some Known Results} 
\label{sec:model}
We use the following notation convention.
Calligraphic symbols denote sets, 
bold symbols denote vectors, 
and  sans-serif symbols denote systematic parameters.
We use $|\cdot|$ to represent the cardinality of a set or the length of a vector;
$[a:b]:=\left\{ a,a+1,\ldots,b\right\}$ and $[n] := [1:n]$; 
$\oplus$ represents bit-wise XOR.  $\arg \max_{x\in \Xc}f(x)=\big\{x\in \Xc:f(x)=\max_{x\in\Xc}f(x)\big\}.$
 In this paper,  we let $\binom{x}{y}=0$ if $x<0$ or $y<0$ or $x<y$.  We use the same convention as that in the literature
when it comes to `summing' sets.

\subsection{System Model}
\label{sub:system model}
Consider the combination network in Fig.~\ref{fig: Combination_Networks}.
The server has access to $\Nsf$ files, denoted by $F_1, \cdots, F_\Nsf$, each composed of $\Bsf$ i.i.d uniformly distributed bits.
The server is connected to $\Hsf$ relays through $\Hsf$ error-free orthogonal links. 
The relays are connected to $\Ksf = \binom{\Hsf}{\rsf}$ users 
through $\rsf \, \Ksf$ error-free orthogonal links. 
%
The set of users connected to  relay $h$ is denoted by $\Uc_{h}, \ h\in[\Hsf]$. 
The set of relays connected to user $k$ is denoted by $\Hc_{k}, k\in[\Ksf]$.
For the network in~Fig.~\ref{fig: Combination_Networks}, for example, 
$\Uc_{1}=\{1,2,3\}$ and 
$\Hc_{1}=\{1,2\}$.

In the placement phase, user $k\in[\Ksf]$ stores information about the $\Nsf$ files in its cache of size $\mathsf{MB}$ bits, where $\Msf \in[0,\Nsf]$.  
The cache content of user $k\in[\Ksf]$ is denoted by $Z_{k}$; let $\Zm:=(Z_{1},\ldots,Z_{\Ksf})$.
During the delivery phase, user $k\in[\Ksf]$ requests file $d_{k}\in[\Nsf]$;
the demand vector $\dv:=(d_{1},\ldots,d_{\Ksf})$ is revealed to all nodes. 
Given $(\dv,\Zm)$, the server sends a message $X_{h}$ 
of $\Bsf \, \Rsf_{h}(\dv,\Zm)$ bits to relay $h\in [\Hsf]$. 
Then, relay $h\in [\Hsf]$ transmits a message $X_{h\to k}$ 
of $\Bsf \, \Rsf_{h\to k}(\dv,\Zm)$ bits to user $k \in \Uc_h$. 
User $k\in[\Ksf]$ must recover its desired file $F_{d_{k}}$ from $Z_{k}$ and $(X_{h\to k} : h\in \Hc_k)$ with high probability when $\Bsf\to \infty$. 
The objective is to determine the optimal {\it max-link load} defined as 
\begin{align}
\Rsf^{\star}(\Msf)
:=
\min_{\substack{\Zm}}\negmedspace
\max_{\substack{k\in\Uc_h, h\in[\Hsf],\\ \dv\in[\Nsf]^{\Ksf}}} \negmedspace
\max 
\left\{
\Rsf_h(\dv,\Zm),
\Rsf_{h\to k}(\dv,\Zm)
\right\}.
\end{align}
The max link-load under the constraint of uncoded cache placement is denoted by $\Rsf^{\star}_{\mathrm{u}}$.


\subsection{MAN Caching Schemes in Shared-Link Networks}
\label{sub:known results}
Since we will use cMAN placement in our proposed delivery scheme, 
we summarize here some known results 
for the shared-link model (where all the $\Ksf$ users are directly connected to the server through an error-free shared-link). 
For the sake of space, we only consider the case $\Ksf \leq \Nsf$.

\paragraph*{Centralized Cache-aided Systems}
Let  $\Msf=t\frac{\Nsf}{\Ksf}$ for some positive integer $t\in[0:\Ksf]$. 
In the {cMAN} placement phase, each file is divided into $\binom{\Ksf}{t}$ non-overlapping sub-files
of length $\frac{\Bsf}{\binom{\Ksf}{t}}$ bits. The sub-files of $F_{i}$ are denoted by $F_{i,\Wc}$
for $\Wc\subseteq[\Ksf]$ where $|\Wc|=t$. 
User $k\in [\Ksf]$ 
fills its cache as
\begin{align}
Z_k 
= \Big( 
F_{i,\Wc}
:  k\in\Wc, 
\ \Wc\subseteq[\Ksf], 
\ |\Wc|=t,
\ i\in[\Nsf]
\Big).
\label{eq:cMAN cache function}
\end{align}
In the delivery phase with demand vector $\mathbf{d}$, for each  $\Jc\subseteq [\Ksf]$ where $|\Jc|=t+1$, 
the multicast messages
\begin{align}
W_{\Jc} := \oplus_{k\in\Jc}F_{d_{k},\Jc\backslash\{k\}},
\label{eq:cMAN multicast messages}
\end{align}
are generated---since user $k\in\Jc$ wants $F_{d_{k},\Jc\backslash\{k\}}$ and knows $F_{d_{j},\Jc\backslash\{j\}}$ 
for all $j\in\Jc\backslash\{k\}$, 
$F_{d_{k},\Jc\backslash\{k\}}$ can be successfully recovered from $W_{\Jc}$.
The achieved load for the shared-link model is  $\frac{\Ksf+t}{1+t}$.
The load can be further reduced when $\Ksf > \Nsf$~\cite{exactrateuncoded}.

\paragraph*{Decentralized Cache-aided Systems}
In the {dMAN} placement phase, each user caches a subset of $\Msf\Bsf/\Nsf$ bits of each file, chosen uniformly and independently at random. 
Given the cache content of users, the bits of the files are naturally grouped into sub-files $F_{i,\Wc}$, where $F_{i,\Wc}$ is the set of bits of file $i\in[\Nsf]$ that are only cached  by the users in $\Wc\subseteq[\Ksf]$.  When $\Bsf\to\infty$ it can be shown that~\cite{decentralizedcoded} 
\begin{align}
\frac{|F_{i,\Wc}|}{\Bsf}\to & \left(\frac{\Msf}{\Nsf}\right)^{|\Wc|}\left(1-\frac{\Msf}{\Nsf}\right)^{\Ksf-|\Wc|}
\textrm{in probability}.
\label{eq:law of large number}
\end{align}
In the delivery phase, for each $t^\prime\in [0:\Ksf-1]$, 
all the $\binom{\Ksf}{t^\prime+1}$ sub-files $(F_{i,\Wc} : |\Wc|=t^\prime, \ i\in[\Nsf])$ are gathered together; since they all 
have approximately the same normalized length as in~\eqref{eq:law of large number}, the 
{cMAN} delivery phase is used for $\Msf=t^\prime \Nsf/\Ksf$ to deliver them. The achieved load for the shared-link model is $\big(\frac{\Nsf}{\Msf}-1 \big) \big[ 1-\big(1-\frac{\Msf}{\Nsf}\big)^{\Ksf}\big]$.
The load can be further reduced when $\Ksf > \Nsf$~\cite{exactrateuncoded}.

\subsection{Bit-Borrowing}
\label{sub:Bit-Borrowing Step}
To conclude this section, we introduce the {\it bit-borrowing idea} proposed in~\cite{deliveryscheme,wan2017finite},  which we will also use in our novel proposed scheme.
For decentralized shared-link caching problems with non-uniform demands or finite file size $\Bsf$, the sub-files in~\eqref{eq:cMAN multicast messages} may have different lengths, as~\eqref{eq:law of large number} may not hold.
If this is the case, instead of zero-padding the sub-files to meet the length of the longest one,
which leads to inefficient transmissions, we can borrow bits from some sub-files to `lengthen' short sub-files in such a way that the borrowed bits need not to be transmitted at a later stage.  
More precisely, if $|F_{d_{\ell},\Jc\backslash\{\ell\}}|<\max_{k\in \Jc}|F_{d_{k},\Jc\backslash\{k\}}|$, 
we take bits from some $F_{d_{\ell},\Wc}$, where $\ell\notin \Wc$ and $\Jc\backslash\{\ell\} \subset \Wc$ (because $F_{d_{\ell},\Wc}$ is also demanded by user $\ell$ and known by the users in $\Jc\backslash\{\ell\}$) and add those bits to $F_{d_{\ell},\Jc\backslash\{\ell\}}$. 

\section{Novel Achievable Delivery Scheme} 
\label{sec:novel scheme}
For combination networks with {cMAN} placement, the schemes in~\cite{cachingincom,multiserver,novelwan2017} first create multicast messages as in~\eqref{eq:cMAN multicast messages} and then 
deliver them to the users by various methods; for example, in~\cite{novelwan2017} one approach for $t=1$ is to use network coding to achieve interference elimination.
In this section we propose a delivery scheme, referred to as {\it Separate Relay Decoding delivery Scheme} (SRDS), based on a novel way to create multicast messages: by leveraging the symmetries in the topology of combination networks, each multicast message sent to relay $h\in[\Hsf]$ is such that it is useful for the largest possible subset of $\Hc_h$ (i.e, users connected to relay $h$).
We highlight key novelties by way of an example.

\subsection{Example for $\rsf=2$}
\label{sub:example of SRDS}
Consider the network in Fig.~\ref{fig: Combination_Networks} with $\Nsf=\Ksf=6$ and $\Msf=t=2$.
With {cMAN} placement, each file $F_{i}$ is partitioned into $\binom{\Ksf}{t}=15$ sub-files of length $\Bsf/15$ bits. 
Let $\mathbf{d}=(1:6)$. 


\paragraph*{Step~1 [Subfile partition]} 
For each $F_{d_{k},\Wc}, k\in[\Ksf]$ where $k\notin \Wc$, we seek to find  the  set of relays in $\Hc_{k}$ each of which is connected to the largest number of users in $\Wc$. Consider the following examples.

For sub-file $F_{1,\{2,3\}}$, which is demanded by user $k=1$ and cached by the users in $\Wc=\{2,3\}$, 
we have that $\Uc_1=\{1,2,3\}$ (relay~1 is connected to two users in $\Wc$) and $\Uc_2=\{1,4,5\}$ (relay~2 is not connected to any user in $\Wc$). So we solve  $\Sc_{k,\Wc}:=\arg\max_{h\in \Hc_k}|\Uc_h \cap \Wc|=\{1\}$ (i.e., relay $h=1$). 
Since $|\Sc_{k,\Wc}|=1$ we simply add $F_{1,\{2,3\}}$ to the set $\Tc^{h}_{k,\Wc\cap\Uc_h}=\Tc^{1}_{1,\{2,3\}}$ representing the set of bits needed to be recovered by user $k=1$ (first entry in the subscript) from relay $h\in\Sc_{k,\Wc}$ (superscript) and already known by the users in $\Wc \cap \Uc_h=\{2,3\}$ (second entry in the subscript) who are also connected to relay $h=1$ (superscript).

Consider now $F_{1,\{2,5\}}$ where $k=1$ and $\Wc=\{2,5\}$. Sub-file $F_{1,\{2,5\}}$ is also demanded by user~1, who is connected to relays $\Hc_{1}=\{1,2\}$.  
Relay~1 is  connected to users $\Wc \cap \Uc_1 = \{2,5\} \cap \{1,2,3\} = \{2\}$, while 
relay~2 is  connected to users $\Wc \cap \Uc_2 = \{2,5\} \cap \{1,4,5\} = \{5\}$, and 
thus we have $\Sc_{k,\Wc}=\arg\max_{h\in \Hc_k}|\Uc_h \cap \Wc|=\{1,2\}$.
Since now $|\Sc_{k,\Wc}|=2$,
we divide $F_{1,\{2,5\}}$ into $|\Sc_{k,\Wc}|=2$ equal-length pieces and denote $F_{1,\{2,5\}}=(F_{1,\{2,5\},h}^{|\Sc_{k,\Wc}|} : h\in\Sc_{k,\Wc})$. We add $F_{1,\{2,5\},1}^{2}$ (i.e., $(k,\Wc,h)=(1,\{2,3\},1)$) to the set $\Tc^{h}_{k,\Wc\cap\Uc_h} = \Tc^{1}_{1,\{2\}}$, and $F_{1,\{2,5\},2}^{2}$ (i.e., $(k,\Wc,h)=(1,\{2,3\},2)$) in the set $\Tc^{h}_{k,\Wc\cap\Uc_h} = \Tc^{2}_{1,\{5\}}$.

After considering all the sub-files demanded by all the users, for relay $h=1$ (and similarly for all other relays) we have
\begin{align*}
 \Tc^{1}_{1,\{2,3\}} &=\{F_{1,\{2,3\}}\}, \   |\Tc^{1}_{1,\{2,3\}}| = \Bsf/15,\\
 \Tc^{1}_{1,\{2\}}   &=\{F^{2}_{1,\{2,4\},1},F^{2}_{1,\{2,5\},1},F_{1,\{2,6\}}\}, \   |\Tc^{1}_{1,\{2\}}| = 2\Bsf/15, \\ 
 \Tc^{1}_{1,\{3\}}   &=\{F^{2}_{1,\{3,4\},1},F^{2}_{1,\{3,5\},1},F_{1,\{3,6\}}\}, \   |\Tc^{1}_{1,\{3\}}| = 2\Bsf/15, \\
 \Tc^{1}_{2,\{1,3\}} &=\{F_{2,\{1,3\}}\}, \   |\Tc^{1}_{1,\{1,3\}}| = \Bsf/15,\\
 \Tc^{1}_{2,\{1\}}   &=\{F^{2}_{2,\{1,4\},1},F_{2,\{1,5\}},F^{2}_{2,\{1,6\},1}\}, \   |\Tc^{1}_{2,\{1\}}| = 2\Bsf/15, \\
 \Tc^{1}_{2,\{3\}}   &=\{F^{2}_{2,\{3,4\},1},F_{2,\{3,5\}},F^{2}_{2,\{3,6\},1}\}, \   |\Tc^{1}_{2,\{3\}}| = 2\Bsf/15, \\
 \Tc^{1}_{3,\{1,2\}} &=\{F_{3,\{1,2\}}\}, \   |\Tc^{1}_{1,\{1,2\}}| = \Bsf/15, \\
 \Tc^{1}_{3,\{1\}}   &=\{F_{3,\{1,4\}},F^{2}_{3,\{1,5\},1},F^{2}_{3,\{1,6\},1}\}, \   |\Tc^{1}_{3,\{1\}}| = 2\Bsf/15, \\
 \Tc^{1}_{3,\{2\}}   &=\{F_{3,\{2,4\}},F^{2}_{3,\{2,5\},1},F^{2}_{3,\{2,6\},1}\}, \  |\Tc^{1}_{3,\{2\}}| = 2\Bsf/15. 
\end{align*}

\paragraph*{Step~2 [Multicast Message Generation]} 
In this example, for each relay $h\in[\Hsf]$ and each $\Jc\subseteq \Uc_h$, we have $\min_{k\in \Jc}|\Tc^{h}_{k,\Jc\setminus \{k\}}|=\max_{k\in \Jc}|\Tc^{h}_{k,\Jc\setminus \{k\}}|$, which is only a function of $|\Jc|$. 
We thus create the multicast messages 
\begin{align}
W_{\Jc}^{h} := \underset{k\in\Jc}{\oplus}\Tc^{h}_{k,\Jc\setminus \{k\}}.
\label{eq:SRDS multicast message}
\end{align}
For example, the server transmits the follows to relay~1, 
\begin{align*}
  & W_{\{1,2,3\}}^{1} = \Tc^{1}_{1,\{2,3\}}\oplus \Tc^{1}_{2,\{1,3\}} \oplus \Tc^{1}_{3,\{1,2\}} 
  \text{of length  $\Bsf/15$ bits}, 
\\& W_{\{1,2\}}^{1} = \Tc^{1}_{1,\{2\}}\oplus \Tc^{1}_{2,\{1\}}
  \text{of length  $2\Bsf/15$ bits}, 
\\& W_{\{1,3\}}^{1} = \Tc^{1}_{1,\{3\}}\oplus \Tc^{1}_{3,\{1\}}
  \text{of length  $2\Bsf/15$ bits, and}
\\& W_{\{2,3\}}^{1} = \Tc^{1}_{2,\{3\}}\oplus \Tc^{1}_{3,\{2\}}
  \text{of length  $2\Bsf/15$ bits.}
\end{align*}

\paragraph*{Step~3 [Multicast Message Delivery]} 
Finally, for each relay $h\in[\Hsf]$ and each set $\Jc\subseteq \Uc_h$ where $W_{\Jc}^{h}\neq \emptyset$,  relay $h$ forwards $W_{\Jc}^{h}$ to the users in $k\in \Jc$.

The normalized (by the file length) number of bits sent from the server to each relay is the same, thus the achieved max link-load is $7/15=14/30$.
The max link-loads of the schemes in~\cite{novelwan2017,cachingincom,multiserver,Zewail2017codedcaching} are 
$17/30$, $20/30$, $20/30$ and $15/30$, respectively. So our proposed scheme performs the best.

\subsection{General SRDS Scheme}
\label{sub:general scheme of SRDS}
The key idea in the above example is that for each relay $h\in[\Hsf]$ and each set $\Jc\subseteq \Uc_h$, the length of the message $\Tc^{h}_{k,\Jc\setminus \{k\}}$ only depends on $|\Jc|$.
However, if $\rsf>2$, 
we may have $\min_{k\in \Jc}|\Tc^{h}_{k,\Jc\setminus \{k\}}|<\max_{k\in \Jc}|\Tc^{h}_{k,\Jc\setminus \{k\}}|$. 
In order to `equalize' the lengths of the various parts involved in the linear combinations for the multicast messages, 
we propose to use the bit-borrowing idea described in Section~\ref{sub:Bit-Borrowing Step}. 
The pseudo code of the proposed SRDS delivery scheme is in Appendix~\ref{sec:pseudo codes}.

\paragraph*{Step~1 [Subfile partition]} 
For each user $k\in [\Ksf]$ and each set  $\Wc\subseteq [\Ksf]\setminus \{k\}$ where $|\Wc|=t$, we search  for the set of relays $\Sc_{k,\Wc}\subseteq \Hc_k$, each relay in which is connected to the largest number of users in $\Wc$,  i.e., $\max_{h\in \Hc_k}|\Uc_h \cap \Wc|$ users. We  partition $F_{d_{k},\Wc}$ into $|\Sc_{k,\Wc}|$ equal-length pieces  $F_{d_{k},\Wc}=(F^{|\Sc_{k,\Wc}|}_{d_{k},\Wc,h} : h\in \Sc_{k,\Wc})$. For each relay $h\in \Sc_{k,\Wc}$, we add $F^{|\Sc_{k,\Wc}|}_{d_{k},\Wc, h}$ to $\Tc^{h}_{k,\Wc\cap\Uc_h}$.

\paragraph*{Step~2 [Multicast Message Generation]} 
Focus on each relay $h\in[\Hsf]$ and each set $\Jc\subseteq \Uc_h$ where $W_{\Jc}^{h}\neq \emptyset$.
For each user $k\in \Jc$, if $|\Tc^{h}_{k,\Jc\setminus \{k\}}|<\max_{k_1\in \Jc}|\Tc^{h}_{k_1,\Jc\setminus \{k_1\}}|$, we use the bit-borrowing step (Step 3-b in Algorithm A) described in Section~\ref{sub:Bit-Borrowing Step}: we 
take bits from $\Tc^{h}_{k,\Wc}$ where $\Jc\setminus \{k\}\subset \Wc$ and $k\notin \Wc$ and add them to $\Tc^{h}_{k,\Jc\setminus \{k\}}$ so that these borrowed bits need not to be transmitted to $k$ later.
Since $\Jc\setminus \{k\}\subset \Wc$, the users in $\Jc\setminus \{k\}$ also knows $\Tc^{h}_{k,\Wc}$. 
After considering all the users in $\Jc$,  the server forms the multicast messages as~\eqref{eq:SRDS multicast message}.

\paragraph*{Step~3 [Multicast Message Delivery]} 
For each relay $h\in[\Hsf]$ and each set $\Jc\subseteq \Uc_h$ where $W_{\Jc}^{h}\neq \emptyset$, the server sends $W_{\Jc}^{h}$ to relay $h$, who then forwards it to each user $k\in \Jc$.

Note that each file is partitioned into at most $\rsf\binom{\Ksf}{t}$ subfiles because each of the $\binom{\Ksf}{t}$ cMAN subfiles is divided into $|\Sc_{k,\Wc}|\leq \rsf$ non-overlapping equal-length pieces.
\subsection{Achievable max link-load for $\rsf=2$}
In Appendix~\ref{sec:proof of load of SRDS} we show that, when $\rsf=2$
the bit-borrowing step is not needed (as in the example in Section~\ref{sub:example of SRDS}). In this case the achieved max link-load is given in Theorem~\ref{thm:load of SRDS}. The computation of the achieved max link-load in closed-form for $\rsf>2$ is part of on-going work.


\begin{thm}
\label{thm:load of SRDS}
For combination network with end-user caches with $\rsf=2$ and $t=\Ksf\Msf/\Nsf\in [0:\Ksf]$, the max link-load is
\begin{align}
&\Rsf^{\star}_{\mathrm{u}}\leq \Rsf_{\mathrm{SRDS}} := 
\frac{\Ksf X_{\Ksf,\Hsf}}{\Hsf \binom{\Ksf}{t}}  
\label{eq:load of SRDS}\\
&X_{\Ksf,\Hsf}:= 
\sum_{b_1=0}^{\min\{t,\Hsf-2\}} 
\frac{1}{b_1+1}\binom{\Hsf-2}{b_1}^{2}\binom{\binom{\Hsf-2}{2}}{t-2b_1}+\\
&
\sum_{b_1=0}^{\min\{t,\Hsf-2\}}\nonumber\sum_{b_2=0}^{b_1-1}\frac{2}{b_1+1}\binom{\Hsf-2}{b_1}\binom{\Hsf-2}{b_2}\binom{\binom{\Hsf-2}{2}}{t-b_1-b_2}.
\label{eq:def of X khr}
\end{align}
\end{thm}

\begin{figure}
\centerline{\includegraphics[scale=0.6]{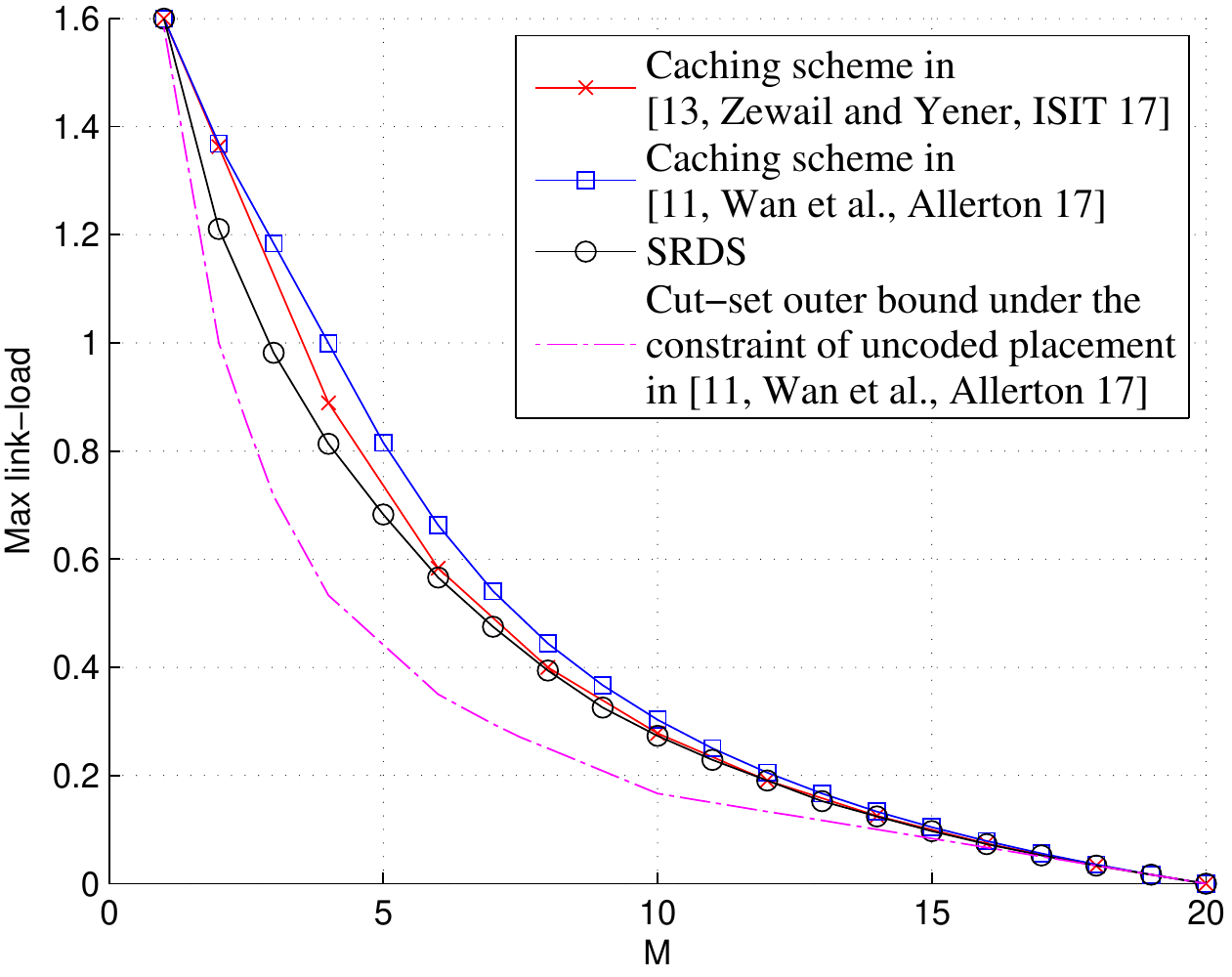}}
\caption{\small Centralized caching system with combination network, where $\Hsf=6$ relays,  $\Nsf=20$ and $\rsf=3$.}
\label{fig:rateh6r3}
\vspace{-5mm}
\end{figure}

\subsection{Numerical Evaluations} 
\label{sub:numerical evaluations}
In this section, we compare the performance of our proposed SRDS with that of existing schemes for centralized combination network for $\Hsf=6$, $\rsf=3$, $\Nsf=\Ksf=20$.  For $\Msf\leq 1$, the load is a 
straight line between $(\Msf,\Rsf)=(0,10/3)$ and
$(\Msf,\Rsf)=(1,8/5)$ achieved by IES in~\cite{novelwan2017}.
%
Fig.~\ref{fig:rateh6r3} shows that our proposed scheme outperforms the schemes in~\cite{novelwan2017} and in~\cite{Zewail2017codedcaching}, which are better than the schemes in~\cite{cachingincom} and in~\cite{multiserver}.

\section{Extensions} 
\label{sec:extension}

In this section, we discuss applications and extensions of our proposed SRDS scheme to other network models
 than centralized combination networks where only end-users have caches. Numerical evaluations  show that our extended SRDS scheme outperforms the state-of-the-art schemes.

\begin{figure}
\centerline{\includegraphics[scale=0.6]{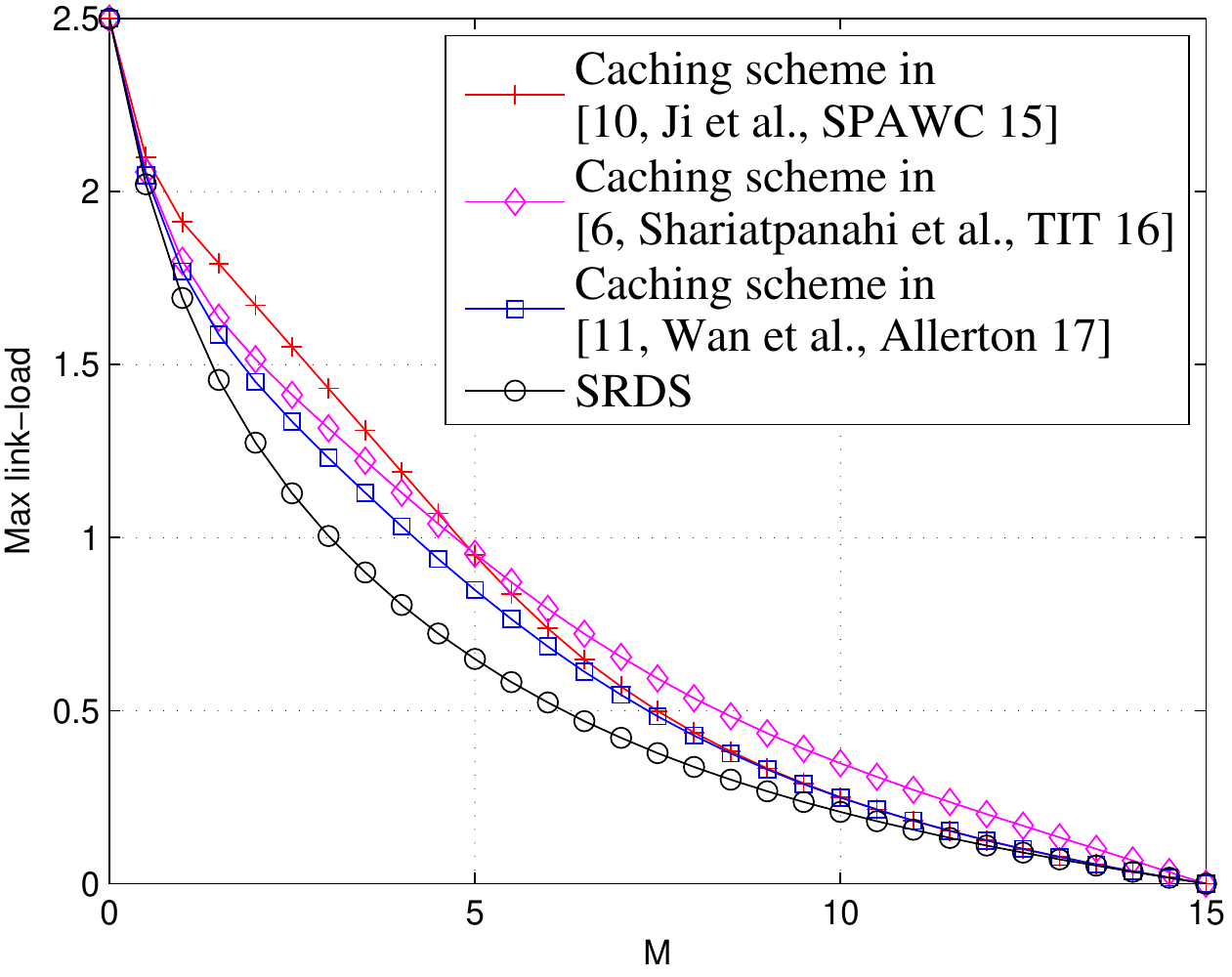}}
\caption{\small Decentralized caching system with combination network, where $\Hsf=6$,  $\Nsf=\Ksf=15$ and $\rsf=2$.}
\label{fig:dech6r2}
\vspace{-5mm}
\end{figure}

\subsection{Decentralized Combination Networks}
\label{sub:decentralized}
Following steps similar to~\cite{decentralizedcoded}, we  extend our proposed delivery scheme
to decentralized combination networks. 
The cache placement phase is the same as {dMAN}. Then for each $t^\prime\in [0:\Ksf-1]$, we gather the sub-files which are only known by $t^\prime$ users and use our proposed delivery scheme to encode those sub-files. More precisely, to encode the sub-files only known by one user, we use the interference elimination scheme proposed in~\cite{novelwan2017}, while to encode the sub-files known by more than one users  we use SRDS.
In Fig.~\ref{fig:dech6r2}, we  compare the performance of the proposed decentralized scheme to those proposed in~\cite{cachingincom,multiserver,novelwan2017} for $\Hsf=6$, $\rsf=2$, $\Nsf=\Ksf=15$. Notice that for each illustrated scheme in Fig.~\ref{fig:dech6r2}, we use the interference elimination scheme proposed in~\cite{novelwan2017} to encode the sub-files known by only one user.
Notice that the placement of the schemes in~\cite{Zewail2017codedcaching,Li2016coded} are designed with the knowledge of the position (the connected relays) of each user in the delivery phase. Hence, it is not possible to extend these schemes to the decentralized case. 
Fig.~\ref{fig:dech6r2} shows the superiority of our proposed scheme over known schemes.

\subsection{General Relay Networks}
\label{sub:general network}
The schemes in~\cite{Zewail2017codedcaching,Li2016coded} can only work for the relay networks where each user is connected to the same number of relays and each relay is connected to the same number of users (we refer to such networks as {\it symmetric networks}). 
In the contrast,
the multicast message generation of SRDS only depends on the connected users of each relay, and thus
can be uses in general relay networks with end-user-caches which are not strictly symmetric. 
In symmetric network, we can directly use SRDS as shown by the next example.

\begin{example}[Symmetric Network]
\label{ex:ex of general relay 1}
Consider a relay network with end-user-caches where $\Nsf=\Ksf=5$, $\Hsf=5$, $\Msf=2$ and
\begin{align*}
&\Uc_{1}=\{1,2,3\},\ \ \Uc_{2}=\{1,3,4\}, \ \ \Uc_{3}=\{1,4,5\},\\
&\Uc_{4}=\{2,4,5\},\ \ \Uc_{5}=\{2,3,5\}.
\end{align*}
Each user is connected to three relays and each relay is connected to three users. 
Let $\mathbf{d}=(1:5)$.
With SRDS, the load to each relay is $1/4$, outperforming the schemes in~\cite{novelwan2017}, in~\cite{Zewail2017codedcaching}, in~\cite{cachingincom}, and in~\cite{multiserver}  whose loads equal $4/15$, $13/45$, $1/3$, and $3/5$, respectively. 
\end{example}

SRDS is designed to minimize the total link-loads to all the relays. 
For symmetric networks, minimizing the total link-loads to all the relays is equivalent to  minimize the max link-load. 
However, for asymmetric networks, we may need to further `balance' the link-load to each relay, which will be shown in the following example.
 
\begin{example}[Asymmetric Network]
\label{ex:ex of general relay 2}
Consider a relay network with end-user-caches where $\Nsf=\Ksf=5$, $\Hsf=5$, $\Msf=2$ and
\begin{align*}
&\Uc_{1}=\{1,2,3\},\ \ \Uc_{2}=\{1,3,4\}, \ \ \Uc_{3}=\{1,4,5\},\\ 
&\Uc_{4}=\{3,4,5\}, \ \ \Uc_{5}=\{2,3,5\},
\end{align*}
i.e., we changed $\Uc_{4}$ from $\{2,4,5\}$ in Example~\ref{ex:ex of general relay 1} to $\{3,4,5\}$ so that the number of connected relays to each user is not the same.
Let $\mathbf{d}=(1:5)$.
We use SRDS  
and indicate the multicast messages as  $W^h_{\Jc,L}=\underset{k\in \Jc}{\oplus}\Tc^{h}_{k,\Jc\setminus \{k\}}$, where in this example we added the subscript $L$ to indicate that the message has length $L$ bits.  
Then, we have the server transmit
\begin{align*}
{\tiny
\hspace*{-0.6cm}
\begin{array}{c c c c c c}
& W^1_{\{1,2,3\},\Bsf/10},   & W^1_{\{1,2\},3\Bsf/20},
& W^1_{\{1,3\},  \Bsf/30},   & W^1_{\{2,3\},\Bsf/20} & \textrm{ to relay }1,\\
& W^2_{\{1,3,4\},\Bsf/10},   & W^2_{\{1,3\},\Bsf/30},
& W^2_{\{1,4\},  \Bsf/20},   & W^2_{\{3,4\},\Bsf/10} & \textrm{ to relay }2,\\
& W^3_{\{1,4,5\},\Bsf/10},   & W^3_{\{1,4\},\Bsf/20},
& W^3_{\{1,5\},  \Bsf/12},   & W^3_{\{4,5\},\Bsf/20} & \textrm{ to relay }3,\\
& W^4_{\{3,4,5\},\Bsf/10},   & W^4_{\{3,4\},\Bsf/20},
& W^4_{\{3,5\},  \Bsf/30},   & W^4_{\{4,5\},\Bsf/20} & \textrm{ to relay }4,\\
& W^5_{\{2,3,5\},\Bsf/10},   & W^5_{\{2,3\},\Bsf/20},
& W^5_{\{2,5\}, 3\Bsf/20},   & W^5_{\{3,5\},\Bsf/30} & \textrm{ to relay }5.\\
\end{array}
}
\end{align*}
It can be seen that the link-loads to relay $1$ to $5$ are $1/3$, $7/30$, $17/60$, $7/30$ and $1/3$, respectively. 
So the achieved max link-load is $1/3$; while the achieved max link-loads by the schemes in~\cite{novelwan2017},~\cite{cachingincom} and in~\cite{multiserver} are $1/3$, $1/2$ and $3/5$, respectively. 

One can further improve on SRDS by observing that the link-load to relay $1$ (or relay $5$) is the largest. 
Thus, instead of transmitting $W^1_{\{1,3\},\Bsf/30}$ to relay $1$, we can transmit it to relay $2$ which is also connected to users in $\{1,3\}$. Similarly, instead of transmitting $W^5_{\{3,5\},\Bsf/30}$ to relay $5$, we can transmit it to relay $4$ which is also connected to users in $\{3,5\}$. With this modification, the achieved max link-load is reduced to $3/10$, which is equal to the cut-set outer bound under the constraint of uncoded placement proposed in~\cite{novelwan2017}.
\end{example}

 The observation at the end of Example~\ref{ex:ex of general relay 2} can be translated in an improvement of the Algorithm in Appendix~\ref{sec:pseudo codes} as described in Appendix~\ref{app:new step 5}.


\subsection{Networks with Cache-Aided Relays and Users}
\label{sub:cache aided relays and users}
Combination networks with both cache-aided relays and  users were considered in~\cite{Zewail2017codedcaching}, where each relay can store $\Msf_1\Bsf$ bits and each user can store $\Msf_2\Bsf$ bits, for $(\Msf_1,\Msf_2)\in[0,N]^2$.
The objective is to determine the lower convex envelop of the load (number of transmitted bits in the delivery phase) pairs 
\[
\big(
\max_{h\in[\Hsf]}\Rsf_h(\dv,\Zm), 
\max_{h\in[\Hsf],k\in \Uc_h}\Rsf_{h\to k}(\dv,\Zm)
\big)
\]
for the worst case demands $\dv$ for a given placement $\Zm$. 

We propose a placement phase combining the ideas of the placement in~\cite{Zewail2017codedcaching} and {cMAN}. We divide each file $F_i$ where $i\in[\Nsf]$ into two non-overlapping parts, $F_i^1$ and $F_i^2$ where  $|F_i^1|=\Bsf\min\{ \rsf\Msf_1/\Nsf,1\}$ and $|F_i^2|=\Bsf\max\{1-\rsf\Msf_1/\Nsf,0\}$. In the placement phase, each relay caches $|F_i^1|/\rsf$ random linear combinations of $F_i^1$ for each file $i\in[\Nsf]$. 
Fix two integers $t_3\in[0:\Ksf]$ and $t_4\in[0:\Ksf]$. Each user firstly randomly and independently caches $t_3|F_i^1|/\Ksf$ bits of $F_i^1$ for each $i\in[\Nsf]$. We then divide each $F_i^2$ into $\binom{\Ksf}{t_4}$ non-overlapping equal-length parts, each of which is denoted by $F^2_{i,\Wc}$ where $\Wc\subseteq [\Ksf]$ and $|\Wc|=t_4$. Each user $k$ then caches $F^2_{i,\Wc}$ for each $i\in[\Nsf]$ if $k\in \Wc$. Hence, $\Msf_2=\frac{t_3\min\{ \rsf\Msf_1,\Nsf\}}{\Ksf}+ \max\{1-\rsf\Msf_1/\Nsf,0\}t_4$. 
In the delivery phase, each relay transmits  $\frac{\Bsf-t_3|F^{1}_{i}|/\Ksf}{\rsf}$  random linear combinations of $F_{d_k}^1$ to each user $k\in \Uc_h$, and uses SRDS to let each user $k$ recover $F_{d_k}^2$. 
The following example compares this proposed scheme and the one in~\cite{Zewail2017codedcaching}.

\begin{example}
\label{ex:ex of relay cache}
Consider the network in Fig.~\ref{fig: Combination_Networks} with $\Nsf=6$, $\Msf_1=1$, $\Msf_2=2$ and $\mathbf{d}=(1:6)$. We divide each file into two parts $F_i^1$ and $F_i^2$, where $|F_i^1|=\Bsf/3$ and $|F_i^2|=2\Bsf/3$. From our proposed caching scheme with $t_3=1$ and $t_4=2$, the achieved link-load pair is  $(14/45,1/3)$ 
while the scheme in~\cite{Zewail2017codedcaching} leads to $(1/3,1/3)$.
We thus see that SRDS is able to lower the link-load  from the source to the relays.
\end{example}

\section{Conclusions}
\label{sec:conclusion and future work} 
In this paper we investigated  centralized combination networks with end-user-caches. By leveraging network topology, we proposed a non-separation approach based delivery scheme, referred to as {\it Separate Relay 
Decoding delivery Scheme} (SRDS), with {cMAN} placement.
 We then extended the proposed SRDS to other relevant models, such as decentralized combination networks, general relay networks, combination network with both cache-aided relays and users. Numerical evaluations showed that in all the aforementioned systems, the proposed scheme outperformed the state-of-the-art schemes.

\paragraph*{Acknowledgement}
The work of K. Wan and D. Tuninetti is supported by Labex DigiCosme and in part by NSF~1527059, respectively.

\appendices

\section{Pseudo Code of Algorithm 1: SRDS}
\label{sec:pseudo codes}

\begin{enumerate}
\item \textbf{input:} $F_{i,\Wc}$ where $i\in[\Nsf]$, $\Wc\subseteq[\Ksf]$ and $|\Wc|=t$;  \textbf{initialization:} $t_1=1$; $\Tc^{h}_{k,\Jc}=\emptyset$ for each $h\in [\Hsf]$, $k\in \Uc_{h}$ and $\Jc\subseteq \Uc_{h}\setminus \{k\}$;
\item \textbf{for}  each $k\in [\Ksf]$ and each $\Wc\subseteq [\Ksf]$ where $k\notin \Wc$,
\begin{enumerate}
\item $\Sc_{k,\Wc}=\arg\max_{h\in \Hc_k}|\Wc \cap \Uc_h|$;  divide $F_{d_k,\Wc}$ into $|\Sc_{k,\Wc}|$ non-overlapping parts with equal length, $F_{d_k,\Wc}=(F^{|\Sc_{k,\Wc}|}_{d_k,\Wc,h}:h\in \Sc_{k,\Wc})$; 
\item \textbf{for}  each $h\in \Sc_{k,\Wc}$, pad $F^{|\Sc_{k,\Wc}|}_{d_k,\Wc,h}$ at the end of $\Tc^{h}_{k,\Wc\cap \Uc_h}$; 
\end{enumerate}
\item \textbf{for}  each $h\in [\Hsf]$ and each $\Jc\subseteq \Uc_h$ where $|\Jc|=t_1$,
\begin{enumerate}
\item $m_1=\max_{k\in \Jc}|\Tc^{h}_{k,\Jc\setminus \{k\}}|$;
\item \textbf{for}  each $k\in \Jc$, \textbf{if} $|\Tc^{h}_{k,\Jc\setminus \{k\}}|<m_1$, \textbf{then}
\begin{enumerate}
\item $R_e=m_1-|\Tc^{h}_{k,\Jc\setminus \{k\}}|$; $t_2=|\Jc|$; ($R_e$ represents the number of bits to be borrowed.)
\item $\Dc=\big\{\Wc\subseteq \Uc_h: k\notin \Wc, \Jc\setminus \{k\}\subset \Wc, |\Wc|=t_2, \Tc^{h}_{k,\Wc}\neq \emptyset\big\}$; ($\Dc$ represents the set of bits which can be borrowed.)
\item \textbf{if}  $R_e\geq \sum_{\Wc\in \Dc}|\Tc^{h}_{k,\Wc}|$, \textbf{then} $\Cc=\underset{\Wc\in \Dc}{\cup}\Tc^{h}_{k,\Wc}$; \textbf{else then}
\begin{enumerate}
\item $\Cc=\emptyset$; sort all the sets  $\Wc\in \Dc$ by the length of $\Tc^{h}_{k,\Wc}$ such that we let $\Dc(1)$ represents the set where $|\Tc^{h}_{k,\Dc(1)}|=\max_{\Wc\in \Dc} |\Tc^{h}_{k,\Wc}|$ while $\Dc(|\Dc|)$ represents the set where $|\Tc^{h}_{k,\Dc(|\Dc|)}|=\min_{\Wc\in \Dc} |\Tc^{h}_{k,\Wc}|$; assume $\Tc^{h}_{k,\Dc(|\Dc|+1)}=\emptyset$;
\item $a$ is the minimum number in $\{i\in [2:|\Dc|]:\sum_{j\in[1:i-1]}|\Tc^{h}_{k,\Dc(j)}|-(i-1)|\Tc^{h}_{k,\Dc(i)}|\geq R_e\}\cup \{|\Dc|+1\}$; 
\item \textbf{for} each $i\in [a-1]$, pad the first $|\Tc^{h}_{k,\Dc(i)}|-\frac{\sum_{i\in[a-1]}|\Tc^{h}_{k,\Dc(i)}|-R_e}{a-1}$ bits of $\Tc^{h}_{k,\Dc(i)}$ at the end of $\Cc$;
\end{enumerate}
\item pad the bits in $\Cc$ at the end of $\Tc^{h}_{k,\Jc\setminus \{k\}}$;
\item \textbf{for}  each $\Wc\in \Dc$, update $\Tc^{h}_{k,\Wc}=\Tc^{h}_{k,\Wc}\setminus \Cc$; 
\item $R_e=R_e-|\Cc|$; \textbf{if} $R_e>0$ and $t_2<|\Uc_h|-1$, \textbf{then} $t_2=t_2+1$ and go to step 3-b-ii);
\end{enumerate}
\item let $W^h_{\Jc}=\underset{k\in\Jc}{\oplus}\Tc^{h}_{k,\Jc\setminus \{k\}}$; 
\end{enumerate}
\item \textbf{if} $t_1<|\Uc_h|$, $t_1=t_1+1$ and go to step 3);
\item \textbf{for} each  relay $h$ and each $\Jc\subseteq \Uc_h$ where $W^h_{\Jc}\neq \emptyset$, transmit
$W^h_{\Jc}$ to relay $h$ and 
 relay $h$ transmits $W^h_{\Jc}$ to each user in $\Jc$;
\end{enumerate}



\section{Proof of Theorem~\ref{thm:load of SRDS}}
\label{sec:proof of load of SRDS}
When $\rsf=2$, each user $k\in[\Ksf]$ is connected to two relays and each relay is connected to $\Hsf-1$ users, say $\Hc_k = \{h,h^{\prime}\}$ for which it holds that
$ \Uc_{h}\cap \Uc_{h^{\prime}}=\{k\}$. 
%
We want to compute $|\Tc^{h}_{k,\Jc\setminus \{k\}}|$ for one relay $h\in [\Hsf]$, one user $k\in \Uc_h$ and one set $\Jc\subseteq \Uc_h$. We consider two cases.
%

Case 1.
The number of $\Wc\subseteq[\Ksf]\setminus \{k\}$, where $|\Wc|=t$, $\Uc_{h} \cap \Wc=\Jc\setminus \{k\}$ and $|\Uc_{h^\prime} \cap \Wc|=|\Uc_h \cap \Wc|=|\Jc|-1$, is $\binom{\Ksf\rsf/\Hsf-1}{|\Jc|-1}\binom{\binom{\Hsf-\rsf}{\rsf}}{t-2(|\Jc|-1)}$. Here, $\Ksf\rsf/\Hsf-1= \Hsf-2$ is the number of users connected to relay $h^\prime$ besides $k$, $\binom{\Hsf-\rsf}{\rsf}$ is the number of users which are  connected neither to $h$ nor to $h^\prime$, and  $t-2(|\Jc|-1)$  represents the number of users in $\Wc$ which are not connected to the relays in $\Hc_k$. 
For each of this type of $\Wc$, we divide $F_{d_k,\Wc}$ into two non-overlapping equal-length parts and put one part in $\Tc^{h}_{k,\Jc\setminus \{k\}}$ and the other part in $\Tc^{h^\prime}_{k,\Uc_{h^\prime} \cap \Wc}$.

Case 2.
The number of $\Wc\subseteq[\Ksf]\setminus \{k\}$, where $|\Wc|=t$,  $\Uc_{h} \cap \Wc=\Jc\setminus \{k\}$ and $|\Uc_{h^\prime} \cap \Wc|<|\Uc_h \cap \Wc|=|\Jc|-1$, is  $\sum_{b_2=0}^{|\Jc|-2}\binom{\Ksf\rsf/\Hsf-1}{b_2}\binom{\binom{\Hsf-\rsf}{\rsf}}{t-(|\Jc|-1)-b_2}$.
For this type of $\Wc$,  we put $F_{d_k,\Wc}$ in $\Tc^{h}_{k,\Jc\setminus \{k\}}$. 

Hence, we showed that for each user $k\in \Jc$, $|\Tc^{h}_{k,\Jc\setminus \{k\}}|$ is identical and we need not to use the bit-borrowing step. So we encode each $F_{d_k,\Wc}$ (or each partitioned piece of it ) by a sum including $\max_{h\in [\Hsf]}|\Uc_{h}\cap \Wc|$ sub-files (or partitioned pieces with the same length).
%
Therefore, link-load to all the relays for transmitting $F_{d_k,\Wc}$ is $|F_{d_k,\Wc}|/\max_{h\in [\Hsf]}\Bsf|\Uc_{h}\cap \Wc|$. 
By considering each integer $b_1=|\Jc|-1\in [0:\min\{t,\Ksf\rsf/\Hsf-1\}]$ and that $\Ksf\rsf/\Hsf-1=\Hsf-2$, the max link-load achieved by SRDS is as in~\eqref{eq:load of SRDS}. 


\section{Pseudo Code of Algorithm 2: New Step 5) for Algorithm 1}
\label{app:new step 5}
\begin{enumerate}
\item \textbf{initialization:} $i=1$; Define that $L_h=\sum_{\Jc\subseteq \Uc_h}|W^h_\Jc|$ for each $h\in [\Hsf]$;
\item sort all the relays $h\in [\Hsf]$ by  $L_h$, i.e.,  $C(1)$ represents the relay whose $L_h$ is maximal and $C(\Hsf)$ represents the relay whose $L_h$ is minimal;
\item \textbf{if} there exists a set $\Jc\subseteq \Uc_{C(i)}$ such that  $|W^{C(i)}_{\Jc}|>0$ and there exists a set of relays (denoted by $\Qc$) where $\Qc\subseteq \{C(i+1),\ldots,C(\Hsf)\}$ and each relay $h\in \Qc$ is connected to all the users in $\Jc$,
\textbf{then, }
\begin{enumerate}
\item choose one relay $h\in \Qc$ where $L(h)=\max_{h\in\Qc} L(h)$, and move $\min\{|W^{C(i)}_{\Jc}|,(L_{C(i)}-L_{h})/2\}$ bits from $|W^{C(i)}_{\Jc}|$ to $W^{h}_{\Jc}$;
\item update $L(h)$ for relay $h$ and update $L(C(i))$ for relay $C(i)$;
\end{enumerate}
\textbf{else then, } $i=i+1$; 
\item \textbf{if} $i<\Hsf$, go to Step 2) of Algorithm 2;
\item \textbf{for} each  relay $h$ and  $\Jc\subseteq \Uc_h$ where $W^{h}_{\Jc}\neq \emptyset$, transmit
$W^{h}_{\Jc}$ to relay $h$, which forwards it to users in $\Jc$;
\end{enumerate}


\bibliographystyle{IEEEtran}
\bibliography{IEEEabrv,IEEEexample}

\end{document}

%% file: macros.tex
\long\def\comment#1{}

\newcommand{\dv}{{\mathbf d}}

\newcommand{\Zm}{{\mathbf Z}}

\newcommand{\Cc}{{\mathcal C}}
\newcommand{\Dc}{{\mathcal D}}

\newcommand{\Hc}{{\mathcal H}}

\newcommand{\Jc}{{\mathcal J}}

\newcommand{\Qc}{{\mathcal Q}}

\newcommand{\Sc}{{\mathcal S}}
\newcommand{\Tc}{{\mathcal T}}
\newcommand{\Uc}{{\mathcal U}}
\newcommand{\Wc}{{\mathcal W}}

\newcommand{\Xc}{{\mathcal X}}

\newcommand{\rsf}{{\mathsf r}}

\newcommand{\Bsf}{{\mathsf B}}

\newcommand{\Hsf}{{\mathsf H}}

\newcommand{\Ksf}{{\mathsf K}}

\newcommand{\Msf}{{\mathsf M}}
\newcommand{\Nsf}{{\mathsf N}}

\newcommand{\Rsf}{{\mathsf R}}

\renewcommand{\arg}{{\hbox{arg}}}

\newcommand{\be}{\begin{equation}}
\newcommand{\ee}{\end{equation}}
\newcommand{\bea}{\begin{eqnarray}}
\newcommand{\eea}{\end{eqnarray}}

